\documentstyle[preprint,eqsecnum,aps,epsf]{revtex}

\def\v12{\vartheta_{12}}

\def\ap#1#2#3   {{\em Ann. Phys. (NY)} {\bf#1} (#2) #3}
\def\apj#1#2#3  {{\em Astrophys. J.} {\bf#1} (#2) #3}
\def\apjl#1#2#3 {{\em Astrophys. J. Lett.} {\bf#1} (#2) #3}
\def\app#1#2#3  {{\em Acta. Phys. Pol.} {\bf#1} (#2) #3}
\def\ar#1#2#3   {{\em Ann. Rev. Nucl. Part. Sci.} {\bf#1} (#2) #3}
\def\cpc#1#2#3  {{\em Computer Phys. Comm.} {\bf#1} (#2) #3}
\def\err#1#2#3  {{\it Erratum} {\bf#1} (#2) #3}
\def\ib#1#2#3   {{\it ibid.} {\bf#1} (#2) #3}
\def\jmp#1#2#3  {{\em J. Math. Phys.} {\bf#1} (#2) #3}
\def\ijmp#1#2#3 {{\em Int. J. Mod. Phys.} {\bf#1} (#2) #3}
\def\jetp#1#2#3 {{\em JETP Lett.} {\bf#1} (#2) #3}
\def\jpg#1#2#3  {{\em J. Phys. G.} {\bf#1} (#2) #3}
\def\mpl#1#2#3  {{\em Mod. Phys. Lett.} {\bf#1} (#2) #3}
\def\nat#1#2#3  {{\em Nature (London)} {\bf#1} (#2) #3}
\def\nc#1#2#3   {{\em Nuovo Cim.} {\bf#1} (#2) #3}
\def\nim#1#2#3  {{\em Nucl. Instr. Meth.} {\bf#1} (#2) #3}
\def\np#1#2#3   {{\em Nucl. Phys.} {\bf#1} (#2) #3}
\def\pcps#1#2#3 {{\em Proc. Cam. Phil. Soc.} {\bf#1} (#2) #3}
\def\pl#1#2#3   {{\em Phys. Lett.} {\bf#1} (#2) #3}
\def\prep#1#2#3 {{\em Phys. Rep.} {\bf#1} (#2) #3}
\def\prev#1#2#3 {{\em Phys. Rev.} {\bf#1} (#2) #3}
\def\prl#1#2#3  {{\em Phys. Rev. Lett.} {\bf#1} (#2) #3}
\def\prs#1#2#3  {{\em Proc. Roy. Soc.} {\bf#1} (#2) #3}
\def\ptp#1#2#3  {{\em Prog. Th. Phys.} {\bf#1} (#2) #3}
\def\ps#1#2#3   {{\em Physica Scripta} {\bf#1} (#2) #3}
\def\rmp#1#2#3  {{\em Rev. Mod. Phys.} {\bf#1} (#2) #3}
\def\rpp#1#2#3  {{\em Rep. Prog. Phys.} {\bf#1} (#2) #3}
\def\sjnp#1#2#3 {{\em Sov. J. Nucl. Phys.} {\bf#1} (#2) #3}
\def\spj#1#2#3  {{\em Sov. Phys. JEPT} {\bf#1} (#2) #3}
\def\spu#1#2#3  {{\em Sov. Phys.-Usp.} {\bf#1} (#2) #3}
\def\zp#1#2#3   {{\em Zeit. Phys.} {\bf#1} (#2) #3}
\def\be{\begin{equation}}
\def\ee{\end{equation}}
\def\theta{\vartheta}
\def\th{\vartheta}
\def\th12{\vartheta_{12}}
\def\thc{\vartheta_c}

\def\al{$ \alpha_s $\ }
\def\a{\gamma_0}
\def\et{\tilde{\epsilon}}
\def\ec{\epsilon_c}
\def\ot{\tilde{\omega}}
\def\k1n{(k_1,..,k_n)}

\def\upo{^{(1)}}
\def\upt{^{(2)}}
\def\r2{r^{(2)}}
\def\d2{d^{(2)}}
\def\la{\lambda}
\def\remark#1{}
\def\rem1#1{#1}

 \newcommand{\labl}[1]{\label{#1}}

\newcommand{\ve}{\epsilon}

\begin{document}
\title{CRITICAL ANGLE IN THE PARTON CASCADE}
\author{Wolfgang Ochs}
\address{Max-Planck-Institut f\"ur Physik, Werner Heisenberg Institut, \\
F\"ohringer Ring 6, D 80805 M\"unchen, Germany}
\author{Jacek Wosiek}
\address{Institute of Physics, Jagellonian University,      \\
PL 30-059 Cracow, Reymonta 4, Poland}
\maketitle
\begin{center}

\vskip3cm
{\bf Abstract}
\end{center}
\vspace{1cm}
The angular correlation function of partons in a jet as
derived from perturbative QCD is nonanalytic at a critical angle
which separates a multiparticle regime with scaling properties from a
regime with few particles near the hadronisation scale. Some
phenomenological consequences are discussed.

\vspace*{4cm}
\noindent MPI-Ph/95-98\newline
TPJU-17/95\newline
November, 1995 \newline
hep-ph/9511457

\newpage

 \section{Introduction}

Recent studies of angular correlations of partons inside jets
within pertubative QCD have demonstrated the emergence of scaling
properties at high energies and at sufficiently large relative angles
\cite{DMO}-\cite{BMP}.
In this paper we focus on the complementary kinematic region of
small relative angles within the same theoretical approach of
perturbative QCD. The correlation function becomes nonanalytic
in the  high energy limit
with a discontinuous second derivative
for a critical angle $\vartheta_c$ which
separates two regimes of quite different characteristics.
In the small angle regime the correlation function is determined by the
one-particle energy spectrum of the parent parton from which the detected
partons emerge. In the theory with running $\alpha_s$ the correlation
function in this new regime depends on the nonperturbative transverse
momentum cutoff $Q_0$ in the cascade
whereas in the large angle regime there was no cutoff dependence.
This opens the possibility to test the hypothesis of
``Local Parton Hadron Duality'' (LPHD) \cite{LPHD} near the
boundary $Q_0$.

The occurrence of a critical angle $\vartheta_c$ in the angular
correlation has been noted already in two previous papers.
In our first study on angular correlations \cite{OW1} we found (for fixed
$\alpha_s$)
a relation between 2-particle correlations, and the 1-particle spectrum
for the critical angle
\be
\vartheta_c =\frac{Q_0}{P} \left(\frac{P\Theta}{Q_0}\right)^{1/5}
\label{thcrit}
\ee
and also for correlations in a pair of points symmetric in certain
variables around this critical angle ($\Theta$ and $P$ are the
opening angle and primary momenta of the jet respectively).

In the analysis by Dokshitzer, Marchesini and Oriani \cite{DMO} the same
angle (\ref{thcrit}) was found as limiting the validity of a power
law in the relative angle from below (again for fixed $\alpha_s$).
Moreover they suggested the existence of a correlation function
below $\vartheta_c$ in a nonperturbative model: the rate for the
particles with
relative angle $\th12$ is given in terms of the momentum distribution
of the parents.

We show in the following that perturbative QCD in its regime of
applicability, ie. for transverse momenta $k_T\geq Q_0$,
provides the correlation function
not only above but also below $\thc$.
It can be expressed indeed by the momentum
distribution of the parent parton in some analogy to the above
effective hadronic model \cite{DMO}.
Furthermore we extend the analysis to the interesting case
of running $\alpha_s$. Again a critical angle can be identified. At high
energies it behaves as
\be
    \theta_c=\frac{\Lambda}{P}
    \left(\frac{\ln (P\Theta/\Lambda)}{\lambda}\right)^{\frac{4}{9}\lambda}
\label{thcritra}
\ee
which depends more weakly on the jet energy
than the angle (\ref{thcrit}) for constant $\alpha_s$
($\Lambda$ is the QCD scale and $\lambda=\ln(Q_0/\Lambda)$).
The special role of the critical angle as singular
point becomes obvious from the fact that the correlation function
depends on the cutoff $Q_0$ only below but not above it.

 \section{Emergence of the critical angle in perturbative QCD}
We consider for definitness the correlation density
$\rho^{(2)}(\th12,P,\Theta)$ in the relative angle $\th12$ of two partons
and comment later on multiparton correlations. In the double logarithmic
approximation (DLA) the correlation function obeys the integral equation
\cite{OW2}
\begin{equation}
\rho^{(2)}(\theta_{12},P,\Theta)=d\upt(\theta_{12},P,\Theta)
     + \int_{Q_0/\theta_{12}}^{P}
\frac{dK}{K} \int_{\theta_{12}}^{\Theta} \frac{d\Psi}{\Psi}
  \a^2(K\Psi) \rho\upt(\theta_{12},K,\Psi).
\labl{rho2}
\end{equation}
where the anomalous dimension $\a$ for the evolution of the multiplicity
is given in terms of the running coupling $\a^2=6\alpha_s/\pi=
\beta^2/\ln{(k_T/Q_0)}$ with $\beta^2=12N_c/(11N_c-2N_f)$. The inhomogenous
term $\d2$ is constructed from the product of 1-particle angular
distributions and found as
$d^{(2)}(x)=\rho^{(1)}(x)
\overline{n}(x),\;\; x=P\th12/\Lambda$ and
with the multiplicity $\overline{n}$ in the cone
of half-angle $\th12$ around the primary parton axis. In exponential
accuracy $d^{(2)} \sim \overline{n}^2$.

The leading contribution to the solution of Eq.(\ref{rho2})
was derived in \cite{OW2}\footnote{Present $R(P/K)$ denotes
$\overline{R}(P/K)+\delta(1-P/K)$ of Ref.{\cite{OW2}}.}
\begin{equation}
\rho^{(2)}(\th12,\Theta,P)=
\int_{Q_0/\th12}^{P} \frac{dK}{K}
R
\left({P\over K}, {\Theta\over\th12},
 {K\th12\over\Lambda} \right)
\d2\left({K\th12\over\Lambda}\right).
 \labl{res}
\end{equation}
The resolvent
$
R
\left({P\over K}, {\Theta\over\th12},
 {K\th12\over\Lambda} \right)$ is the momentum distribution of the
parent partons $K$ in the jet $(P,\Theta)$ with the condition, that
their virtuality is bounded from below by $K\th12$ and {\em not} by the
elementary cut-off $Q_0$. This is equivalent to limiting
from below their emission
angle by $\th12$. The representation (\ref{res}), which was derived
as the solution of the differential equation (\ref{rho2}) is identical
with the jet calculus result used in Ref.\cite{DMO}.

 In order to show the existence of the nonanalyticity of the
$\rho^{(2)}(\th12)$ given by
Eq.(\ref{res})  at some small value of the relative angle
$\th12=\thc$,
 consider the dependence of the integrand
of Eq.(\ref{res}) on the parent parton momentum $K$. The interplay
between the decrease of the parent density $R$ and the increase
of the children densities $\d2$ with $K$ produces a sharp maximum
which is however
$\th12$ dependent. The important element
of running \al analysis is that the position of this maximum is {\em
independent of} $Q_0$, since both distributions do not depend on $Q_0$
\footnote{Except of the overall normalization factor which does not
influence the shape of the momentum dependence,
hence is not relevant for the argument.}.
Now, for small relative angles the above maximum shifts {\em below}
the lower limit $Q_0/\th12$ of the momentum integration. At that point
the result changes nonanalytically (albeit very gently - analogously to the
second order phase transitions in statistical systems). Namely, for
$\th12$ bigger than $\thc$, $\rho^{(2)}$ is given ( to the exponential
accuracy) by the $Q_0$ independent value of the integrand at the
maximum. For $\th12$ below $\thc$ however, the result is given by the
value of the integrand {\em at the lower bound} and obviously depends on
$Q_0$. Moreover, since $d_2(1)=1  $ in our exponential approximation,
the density of pairs below
$\thc$ is given entirely by the density of parent partons at the
minimal momentum $K=Q_0/\th12$.

It is easy to verify explicitly the above considerations
for constant \al.
In the logarithmic variables which are natural for this problem
\begin{eqnarray}
Y=\ln{(P\Theta/Q_0)}, &
u=\ln{(P\th12/ Q_0)},& v=\ln{(\Theta/\th12)}, \\
  & \xi=\ln{(P/ K)}=\ln{(1/ x)}, &
\labl{logs}
\end{eqnarray}
the solution (\ref{res}) reads
\begin{equation}
\rho^{(2)}(u,Y)=\int_0^u d\xi R(\xi,v) d_2(u-\xi).  \labl{rcon}
\end{equation}
Using the known momentum distribution of parent partons
$\rho\upo(\xi,Y)=\exp{(2\a\sqrt{\xi(Y-\xi)})}$ and
  replacing the vituality cut-off $Q_0\rightarrow K\th12$, i.e.
$Y\rightarrow v+\xi$,
gives
\begin{equation}
\rho^{(2)}(u,Y)=\int_0^u \exp{\left(\a(2\sqrt{\xi v}+n(u-\xi))\right)} d\xi,
 \labl{in2}
\end{equation}
where we have also substituted the asymptopic form of the inhomogenous term
$d_n(u-\xi)=\exp{(n\a (u-\xi))}$. The integer $n$, which for the two-body
correlations is just $2$, will be kept arbitrary in the following discussion.
This allows for the direct generalization to the  higher order correlations
and multiplicity moments. The maximum of the integrand in (\ref{in2})
is located at
\begin{equation}
\xi_{max}={n^2\over v}. \labl{sad}
\end{equation}
This remains in the integration range only if
\begin{equation}
 \et < {n^2\over n^2+1}, \label{crit}
\end{equation}
where we have introduced the scaling variable
\begin{equation}
\et={v\over u+v}={\ln{(\Theta/\th12)}\over\ln{(P\Theta/Q_0)} }.
\labl{til}
\end{equation}
In this case the result for the density reads
\begin{equation}
\rho^{(2)}(u,Y)\sim\exp{ \left( \a (n u + {v\over n}) \right) },
\end{equation}
and we obtain the power behaviour in the angle $\rho\upt \sim
\th12^{(3\a/2)}$.
If $\xi_{max} > u$ the integral is saturated by the maximal value of the
integrand at the boundary and we obtain
\begin{equation}
\rho^{(2)}(u,Y)\sim R(u,v)\sim \exp{\left(2\a\sqrt{u v}\right)},\;\;\;\;
\et > {n^2\over
n^2+1},
\end{equation}
i.e. the two-parton distribution is given in terms of the single
parton momentum distribution.
One can summarize our results in the "$\epsilon$ - scaling" form
\begin{eqnarray}
\rho^{(2)}(\et,Y)&\sim &\exp{(\a Y \tilde{\omega}(\et) ) }, \labl{oc}\\
\ot(\et)=& n-(n-{1\over n})\et,   & \;\;\;\;\et < \et_c,  \labl{scal1}\\
\ot(\et)=& 2\sqrt{\et(1-\et)}, & \;\;\;\;\et > \et_c.
\labl{escal}
\end{eqnarray}
where $\et_c=n^2/(n^2+1)$. Note that the solution (\ref{escal})
is $n-$independent above $\et_c$.
 The critical angle in Eq. (\ref{thcrit}) quoted in the Introduction
corresponds to  $\et_c=4/5$ for $n=2$.
It is always bigger than the minimal
relative angle $\vartheta_{min}=Q_0/P$ and the
maximal  transverse momentum
associated with $\thc$ grows like $(P/Q_0)^{1/5}$.
The same result can be obtained from the
exact solution
for $\rho\upt$ in terms of Bessel functions \cite{OW1}
which is valid also at finite energies.

The above solutions are displayed in Fig.1 for different $n$.
The linear solutions for $\et < \et_c$
are tangent to the ellipse (\ref{escal});
at the critical point $\et_c$
the functions $\ot(\et)$ and their derivatives are continuous
but not their second derivative.

The same asymptotic result (\ref{oc}) for $\rho\upt$ is also obtained from the
nonlinear, first order differential equation
\begin{equation}
(\ot-\et\ot')(\ot+(1-\et)\ot')=1
\labl{difeq}
\end{equation}
which corresponds to Eq.(\ref{rho2})
 \cite{OW2}.
One finds the linear solution, Eq. (\ref{scal1}), for any boundary value
$\ot(0)=n>0$ and the ellipse solution, Eq.(\ref{escal}),
for the boundary value $\ot(1)=0$. Eq. (\ref{difeq})
can be solved explicitly for $\ot'$ with two roots in general;
for a point on the ellipse there is only one root corresponding to
a bifurcation point where the two solutions with proper boundary conditions
at $\et=0$ and $\et=1$ meet.

The simpler case of fixed \al or $\a$
considered above is instructive as it allows for the exact solution
\cite{OW1}, in this case the dependence on the QCD scale $\Lambda$ in
Eq.(\ref{rho2}) drops out and the only reference scale is provided by the
 cut-off $Q_0$. The constant \al solution is also obtained from the running
 \al result  when both $\beta^2$ and
$\la=\ln{Q_0/\Lambda} \rightarrow\infty$ with their ratio
$\beta^2/\la=\a^2$ fixed. This is equivalent to the famous limit
$N_f\rightarrow 11N_c/2$ with $\Lambda\rightarrow 0$ tuned in such a way that
the anomalous dimension $\a$ remains fixed. In other words $\Lambda$ behaves
as
\begin{equation}
\Lambda=Q_0 \exp{\left( - {12N_c\over \a^2(11N_c-2N_f) } \right) }
\rightarrow 0,
\end{equation}
in this limit.

\section{Running \al and double scaling}

 We employ  the Mellin
representation of the inclusive momentum distribution for running \al ,
see Appendix B of Ref.\cite{OW1} for the details. With the appropriate
asymptotic forms of the energy moments one obtains
\begin{equation}
\rho^{(2)}(\th12,\Theta,P)\approx\int_{\gamma}
\exp{(s v)}
{\cal F}(s, L)
\exp{w_n(s, l)}
{ds\over 2\pi i},
\labl{ds5}
\end{equation}
where
\begin{eqnarray}
{\cal F}(s,x) = & e^{{x\over 2}g_{-}(s,x)} &
\exp{\left[{2\beta^2\over s}\ln{({\sqrt{x}\over 2\beta}g_{+}(s,x)})\right]},
\labl{fas}\\
 \exp{w_n(s,l)}=&
\int_{\lambda}^l d\sigma  &
\exp{v_n(s,\sigma)},
\labl{ds66} \\
v_n(s,\sigma)=&-{\sigma\over 2} g_{-}(s,\sigma)-&
{2\beta^2\over s}\ln{({\sqrt{\sigma}\over 2\beta} g_{+}(s,\sigma))}
+2n\beta\sqrt{\sigma},   \nonumber \\
g_{\pm}(s,x)&=&\sqrt{s^2+4\beta^2/x}\pm s,
\labl{dsigma}
\end{eqnarray}
and all prefactors have been neglected. The logarithmic variables
are now defined with $\Lambda$ as the reference scale
\begin{equation}
L=\ln{(P\Theta/\Lambda)},\;\;\; l=\ln{(P\th12/\Lambda)},\;\;\;
\sigma=\ln{(K\th12/\Lambda)},\;\;\; \lambda=\ln{(Q_0/\Lambda)},
\end{equation}
and $v$ as in Eq.(\ref{logs}).
The saddle point of the $\sigma$ integration is located at
\begin{equation}
\sigma^{\star}={\beta^2(1-n^2)^2\over s^2 n^2}. \labl{sadsig}
\end{equation}
However the maximum of the $\sigma$ integrand is not always within the
bounds of the real $\sigma$ integration. In particular for
\begin{equation}
s=s_{\odot}= -{\beta\over \sqrt{\lambda}}{n^2-1\over n},
\end{equation}
this maximum coincides with the lower cut-off at $\sigma=\lambda$.
Second characteristic point in the complex $s$ plane is the saddle point
$s_{\times}$ of the $s$ integration.
 Depending on the external parameters (e.g. $\th12$) $s_{\times}$
can be bigger or smaller than $s_{\odot}$ thus determining the leading
contribution to the integrand \footnote{The value of $s_{\times}$ is
real.} . The condition that the two
points coincide determines the critical angle $\thc$. In summary:
for all $\th12$ the density of pairs can be written in the form
\begin{equation}
\rho^{(2)}(\epsilon,L,\rho)\sim\exp{\left(2\beta\sqrt{L}(
\omega_n(\epsilon,\rho)-2\rho)\right)}, \labl{ro}
\end{equation}
with
\begin{equation}
\epsilon={v\over L}={\ln{(\Theta/\th12)}\over\ln{(P\Theta/\Lambda)} },
\;\;\;\rho=\sqrt{\lambda/L}.
\end{equation}
For $\th12 > \thc $ both integrations are done by the saddle
point approximation with the result, $z=s\sqrt{L}/\beta$,
\begin{eqnarray}
\omega_n(\epsilon)&=&\gamma(z(\epsilon))+\epsilon
z,\labl{ome}\\ \gamma(z)&=&{1\over 2} (\sqrt{z^2+4}-
z),\labl{gamaofz}
\end{eqnarray}
and $z_n(\epsilon)$ is the
solution of the following algebraic equation
\begin{equation}
\gamma^2(z)-
2\ln{\gamma(z)} - \epsilon z^2 = n^2 - 2\ln{n}. \labl{alg}
\end{equation}
For $\th12 < \thc$, one uses the value of the $\sigma$ integrand
at the lower boundary $\sigma=\lambda$ as the estimate for the
$\sigma$ integration which results in the different saddle point
of the $s$ integration. The final result reads in this case
\begin{eqnarray}
\omega_n(\epsilon,\rho)&=&\gamma(z)+\epsilon
z - \rho (\gamma(z\rho)-n),\labl{omer}
\end{eqnarray}
and $z_n(\epsilon,\rho)$ is determined by the different saddle point condition
\begin{equation}
\gamma^2(z)-
2\ln{\gamma(z)} - \epsilon z^2 = \gamma^2(z\rho)-2\ln{\gamma(z\rho)}.
\labl{algro}
\end{equation}
The critical point
\begin{equation}
z_{\odot}=s_{\odot} \sqrt{L}/\beta=-{n^2-1\over n\rho},
\labl{zc}
\end{equation}
coincides with the saddle point at $\th12=\thc$ or equivalently
at $\epsilon=\epsilon_c$. This fact can be used to determine
$\epsilon_c$ from e.g. Eqs.(\ref{alg}) and (\ref{zc}). The consistency
check of such a procedure is the {\em equivalence } of both equations
(\ref{alg}) and (\ref{algro}) at $z=z_{\odot}$. This is readily verified
since $\gamma(z_{\odot}\rho)=n$. Our result for $\epsilon_c$ reads
\begin{equation}
\epsilon_c(\rho)={1\over z_{\odot}^2}\left(\gamma^2(z_{\odot})-n^2
-\ln{(\gamma^2(z_{\odot})/n^2)} \right).
\labl{ec}
\end{equation}
or explicitly for $n=2$ (see Fig.2)
\begin{equation}
\epsilon_c(\rho)={1\over 4}(\sqrt{1+({4\rho\over 3})^2}+1)^2
-({4\rho\over 3})^2-{1\over 2} ({4\rho\over 3})^2 \ln{\left(
{3\over 4\rho} (\sqrt{1+({4\rho\over 3})^2}+1)/2 \right) }.
\labl{ec2}
\end{equation}
Figure 3 shows the $\omega$ functions obtained from Eqs.(\ref{ome} -
\ref{algro}) for few values of $\rho$ .
As expected from the general discussion in Sect.2,
$\epsilon_c$ and $\omega_n(\epsilon,\rho)$ (for $\epsilon>\epsilon_c$ )
depend on  $Q_0$. This is in contrast to
the constant \al case,
where the only scale was provided by $Q_0$ itself. For the running
\al, the second scale $\Lambda$ allows to construct two dimensionless
ratios relevant for this problem\footnote{In the DLA the limit
$\lambda\to\infty$ would lead to a divergence of the multiplicity and is
therefore not allowed},
 and, as a consequence, the simple
$\epsilon$ scaling is violated for $\epsilon > \epsilon_c$ ($\th12 < \thc$).
On the other hand, $\omega_n(\epsilon,\rho)$ is independent of $n$ for
$\epsilon>\ec(n)$ as in the fixed \al case.

It follows from Eq.(\ref{algro}) that the new solution, Eq.(\ref{omer}),
has the correct behaviour at the kinematical boundary $\th12=Q_0/P$.
One verifies readily that Eq.(\ref{algro}) has a real solution
only for $\epsilon < \epsilon_{max}=1-\rho^2$  which corresponds to the
above kinematic limit for $\th12$. Around that point $\omega$ has the
following expansion
\begin{equation}
\omega(\epsilon,\rho)\simeq 2\rho
-2\ln{(\rho)}\sqrt{(\epsilon_{max}-\epsilon)}
+A_3\sqrt{(\epsilon_{max}-\epsilon)}^3+ ...
\end{equation}
with
\begin{equation}
A_3={
   2\rho^2\ln^{3/2}{(1/\rho^2)}-2\rho^2\ln{(1/\rho^2)}-(1-\rho^2)(1+2\rho^2)
  \over
   2\rho^2\ln^{3/2}{(1/\rho^2)}
    }.
\end{equation}
This is to be contrasted with the asymptotic behaviour of the infinite
energy limit $\omega(\epsilon)$ at $\epsilon \sim 1$ obtained from
Eqs.(\ref{ome}-\ref{alg}).
\begin{equation}
\omega(\epsilon)\simeq \sqrt{(1-\epsilon)\ln{1\over 1-\epsilon}}.
\labl{omas}
\end{equation}
The finite value of $\omega(\epsilon,\rho)=2\rho$ at the boundary
ensures our proper high energy normalization of the density
 $\rho^{(2)}\sim 1$
at $\th12=Q_0/P$. The
 coefficient $\ln{(1/\rho^2)}$ of the threshold behaviour
turns into the additional logarithmic singularity of the asymptotic solution
(\ref{omas}) around $\epsilon \sim 1$.

The constant \al results of the previous Section can be recovered from
Eq.(\ref{ec}) in the limit $\lambda\rightarrow\infty$ or $\rho\rightarrow 1$.
One finds
\begin{equation}
\epsilon_c \simeq {4\over 5} (1-\rho^2), \;\;\;\;\; \rho\rightarrow 1,
\end{equation}
which agrees with (\ref{escal}) since the scaling variable for the constant
\al,
$\tilde{\epsilon}=\epsilon/(1-\rho^2)$.

On the other hand in the high energy limit, $L\rightarrow\infty, Q_0 -
{\rm fixed}$, $\rho\rightarrow 0$ and Eq.(\ref{ec2}) gives
\begin{equation}
\epsilon_c \simeq 1-{4\rho^2\over 9}\ln({1\over\rho^2}), \;\;\;\;
\rho\rightarrow 0, \labl{ecinfty}
\end{equation}
and the critical angle
\begin{equation}
\thc={\Lambda\over P} \exp{(L(1-\epsilon_c(\rho)))},  \labl{thcro}
\end{equation}
may become small raising the question of the applicability of
perturbative QCD in the new regime ($\th12 < \thc$).
However it is easy to show that the transverse momentum $P\thc$ is always
bigger than the perturbative cut-off $Q_0$ which can be written as
\begin{equation}
Q_0=\Lambda \exp{(L\rho^2)}.\labl{q0}
\end{equation}
Since
\begin{equation}
\epsilon_c(\rho) < \epsilon_{max}=1 - \rho^2   \labl{ineq}
\end{equation}
is always in the physical region ( $\thc > \th12^{min}=Q_0/P$ ) ,
  perturbative QCD  applies also for $ Q_0/P < \th12 < \thc $.
The above argument is valid
only for $\rho > 0$. Exactly at $\rho=0$ (infinite energy, at fixed $Q_0$)
the inequality (\ref{ineq}) is saturated and using the asymptotic
form (\ref{ecinfty}) we obtain result (\ref{thcritra}) quoted in the
Introduction.
It implies the transverse momentum associated with $\thc$
 grows as a power of the logarithm of $P$ and is therefore
larger than $Q_0$.

At the critical angle the $\omega$ function, hence also $\rho^{(2)}$,
are known analytically since the solution of the
Eqs.(\ref{alg},\ref{algro}),
 i.e. $z_{\odot}$, is known explicitly, c.f. Eq.(\ref{zc}). In particular
in the high energy limit, Eq.(\ref{ecinfty}), we obtain
\begin{equation}
\rho^{(2)}(\thc,P,\Theta)=
\left({\ln(P\Theta/\Lambda)\over\lambda}\right)^{3\beta\sqrt{\lambda}},
\end{equation}
i.e. the density of pairs with the critical opening angle is growing
only with a power of the logarithm of the energy.

The transition between the high
energy and constant \al limits can be conveniently presented in
slightly different variables. In Fig.4 we show the function
$\omega_R(\et,\rho)$
\be
\omega_R(\et,\rho)=(\omega(\et(1-\rho^2),\rho)-2\rho)/(1-\rho), \labl{omr}
\ee
 as the function of $\et$,
in the normalized interval $0<\et< 1$, and for several values of $\rho$.
The constant \al regime (c.f. Fig.1 ) is already reached for $\rho \sim .8$

The limiting behaviour, Eq.(\ref{ecinfty}) implies that the whole region
$\th12<\thc$ shrinks to zero ($ \epsilon = 1 $) at infinite energy
and constant $Q_0$. Therefore the existence of the non-scaling part
$\omega(\epsilon,\rho)$ can be regarded as the non-leading effect. However
there is an important distinction between this type of finite energy
corrections and the behaviour for $ \th12 > \thc$, namely
 below $\thc$ the leading contribution  is $Q_0$
{\em dependent}, while above $\thc$ it is not. This follows from the
general argument presented in Section 2. Second point is that precisely
these corrections lead to the  finite value of $\tilde{\epsilon}_c$
in the constant \al limit. In other words, the high energy and constant
\al limits are not interchangeable at fixed $Q_0$, and the region
$\th12 < \thc$
is responsible for the difference. With that point in mind, one can define
the double scaling limit:
\begin{equation}
L \rightarrow\infty,\;\;\; \epsilon,\;\;\rho - {\rm fixed}. \labl{ds}
\end{equation}
In that limit $\omega(\epsilon,\rho)$ is the true asymptotic scaling
function of two scaling variables, for all $0<\epsilon,\rho<1$. It
 is singular at
$\epsilon=\epsilon_c(\rho)$ and becomes independent of $\rho$ for
$\epsilon<\epsilon_c(\rho)$.
The double scaling limit (\ref{ds}) requires $Q_0$ to grow with
energy ( $L$ ) according to Eq.(\ref{q0}).
This is not quite along the standard assumption of constant $Q_0$ at high
energies. However this limit allows us to make precise prediction in the
$L\rightarrow\infty$ limit for the two regions above and below $\ec$
which are expected to persist for finite energies.

More importantly,
comparison of Eqs.(\ref{thcro}) and (\ref{q0}) shows that the necessary
condition for applicability of the perturbative QCD, is now safely satisfied
since the $P\thc$ grows exponentially faster than $Q_0$ due to (\ref{ineq}).
Therefore even the softer parents are described by the perturbative
approach in this limit. The singularity nature of the $\omega$ function at
$\epsilon_c(\rho)$ follows readily from Eqs.(\ref{ome} - \ref{algro}):
$\omega(\epsilon,\rho)$ is continuous together with its $\epsilon$ derivative
but the second derivative is not.

 By changing $\rho$ within $0< \rho <1$ one can move the bifurcation point
$\epsilon_c$ through the whole range of $0<\epsilon<1$. Since at $\epsilon_c$
the $\omega$ function is known analytically and, moreover, the solution of the
algebraic equations (\ref{alg},\ref{algro}) is also known
(c.f. Eq.(\ref{zc})) we can provide the explicit analytic
form for the scaling function in the parametric representation for all
$\epsilon$
\begin{equation}
\epsilon(t)={1\over 4}(\sqrt{1+t^2}+1)^2-t^2 -{t^2\over 2}
\ln{\left({1\over 2t}
(\sqrt{1+t^2}+1)\right)}
\end{equation}
\begin{equation}
\omega(t)=-{2\over t}\epsilon(t)+{1\over t}(\sqrt{1+t^2}+1)
,\;\;\;    0<t<{4\over 3}.  \labl{param}
\end{equation}
For $0<t<4/3$ the asymptotic curve $\omega(\epsilon,\rho=0)$ is produced.
For finite  non-zero $\rho$ the representation (\ref{param}) is valid for
  $4\rho/3 < t < 4/3$ and describes  the $\rho$ independent branch
of the scaling function  $\omega_n(\epsilon)$ for
$\epsilon < \epsilon_c$.

\section{Conclusions and phenomenological applications.}
The angular correlation functions exhibit two kinematic regimes at
high energies with quite different properties, separated by a critical
angle $\thc$ or $\ve_c$. Whereas for fixed $\alpha_s$ this critical
behaviour occurs for fixed energy independent $\ve_c$, in the realistic
case of running $\alpha_s$ we find $\ve_c\to 1$ for $P\to \infty$.
We consider here the preasymptotic region with $\ve_c<\ve_{max}<1$
(for $Q_o/\Lambda>1$) where besides the leading terms of the order
$\ln \rho^{(2)}\sim\sqrt{L}$, the terms of all orders
in  $\sqrt{\lambda/L}$ are kept
($\lambda=\ln(Q_o/\Lambda)$, $L=\ln(P\Theta/\Lambda)$).
Then the boundary conditions with finite $\ve_{max}$ can be fulfilled.
The occurence of the critical value $\ve_c$ is quite special to angular
correlations and, for example, there is no such
critical value in the distribution of the energy variable
$\xi=\ln 1/x$ (the ''hump backed plateau``): so, in case of fixed $\alpha_s$,
the function $\ln\rho^{(1)}(\xi)$ is asymptotically given by
$\sqrt{\xi(Y-\xi)}$ in the full kinematic region
whereas the angular correlation function is built up from
the two different pieces in Eq.(\ref{escal}) with discontinuous
second derivative.

In the following we want to point out two differences between
the two regimes (in the theory with running $\alpha_s$) with
their phenomenological consequences.

\noindent 1. Dependence of the cutoff $Q_0$.

\noindent In the region $\ve<\ve_c$
the correlation function is $Q_0$-independent and develops
a scaling behaviour ($\ve$-scaling \cite{OW2}) whereas in the
complementary region ($\epsilon > \ec$) this scaling is broken by $Q_0)$.
This leads to two
possible effects:\\
\noindent a) as a function of energy the small $\ve$ region approaches
the scaling limit which depends on the parameter $\Lambda$,
but not on $Q_0$, whereas at
large $\ve$ the correlation function drops towards $\ve_{max}$ which depends
in addition
on the cutoff $Q_0$ with $\ve_{max}\to 1$
for $P\to\infty$, so it is nonscaling.
This  opens the possibility to determine independently the parameter
$\Lambda$ from the small $\ve$ data and the
effective cutoff $Q_0$
from the large $\epsilon$ data. The first results from DELPHI \cite{DELPHI}
on ``$\ve$-scaling'' could actually be interpreted along these lines:
data with smaller $P\Theta$ drop down earlier with increasing $\epsilon$.
For more quantitative results an analysis with higher accuracy seems to be
required: on the theoretical side an analysis beyond the DLA
would be desirable and on the
experimental side a better treatment of the jet axis problem, for example,
by using the Energy-Multiplicity-Multiplicity observable\cite{EMMC,OW3};
\\
\noindent b) another effect is the dependence of the correlations
on the species of particles under the assumption that
the
particle mass acts as an effective transverse momentum cutoff
and is therefore related to $Q_0$ \cite{LPHD}.
The correlation functions for particles of different species
(say $\pi \pi,~KK, ~pp$) should then
be the same for small $\ve$ and different for large $\ve$ where
$\rho^{(2)}$ for heavier particles vanishes for smaller $\ve_{max}$.

\noindent 2. Dependence of the correlation functions
on their order $n$.    \\
 A convenient measure of correlations of higher order are the
multiplicity moments, for example the factorial moments
$f^{(q)}=<n(n-1)...(n-q+1)>$ in a limited angular
ring (dimension $D=1$) or cone ($D=2$) of half opening
$\delta$ at polar angle $\theta$ to the primary parton (jet) direction
\cite{OW2,DD,BMP}. These moments (usually one considers the normalised moments
 $F^{(q)}=f^{(q)}/<n>^q$) behave at high energies like\footnote{
Factorial $F^{(n)}$ or cumulant $C^{(n)}$ moments approach the same
asymptotic limit but the nonleading corrections appear to be smaller for the
$F^{(n)}$ \cite{OW2,DELPHI}.}
\be
 \ln{F}^{(n)} \sim 2\beta\sqrt{L} \omega_n(\ve,\rho).
\label{lnfq}
\ee
One can determine the function $\omega_n(\ve,\rho)$
 from the measurement
of $F_n$
after  appropriate rescaling \cite{OW2} (``$\omega_n^F$'').
 Again for smaller $\ve$ one should obtain the
predicted
scaling limit $\omega_n(\epsilon,0)$
 whereas at sufficiently large $\epsilon>\epsilon_c(n)$
 there is the dependence
on $\rho=\sqrt{\la/L} $ but no dependence on the order $n$.
This prediction follows as the opening angle $\delta$ determines
the parent momentum $K>Q_0/\delta$ independent of the number of particles
inside the cone because $d^{(n)}=1$ at the threshold independently of $n$.
In the same way one can extract $\omega_2^{\rho}(\epsilon,\rho)$ from
$\rho^{(2)}$.
Then all rescaled correlations
 should yield the same scaling
function $\omega(\epsilon,\rho)$ independent of $n$ for large $\epsilon$
\be
\omega_n^F\sim \omega_2^{\rho}.
\label{frel}
\ee
Also in this region (large $\epsilon$) the cutoff $Q_0$ could
possibly be determined as lower
limit of $K\delta$ or $K\vartheta_{12}$ where $K$ is the momentum of the
$n-$cluster or the pair considered.

We have discussed here the consequences of perturbative QCD in a
preasymptotic region at high energies. At the energies realistic today one has
to consider also finite energy effects. At high energies the width
$\sigma^2$ of the
peak in the variable  $\xi/Y$ of the integrand in Eqs. (\ref{res},\ref{rcon})
 decreases like $1/\sqrt{L}$
and the two regimes become separated at $\ve_c$. At finite energies
there is an overlap between these regimes of this order.
To illustrate the finite energy effects we show in Fig. 5   the rescaled
normalized correlation function with the asymptotic behaviour \cite{OW2}
\be
\hat{r}(\epsilon)\equiv\frac{\ln r(\v12)}{\sqrt{\ln P\Theta/\Lambda}} =
  2 \beta (\omega(\ve,2) - 2\sqrt{1-\ve})
\label{lnrsc}
\ee
and the exact numerical solution of the integral equation (\ref{rho2})
for $\rho^{(2)}(\v12) $ at LEP energies for different opening angles
$\Theta$. One can see that the scaling behaviour is approached quickly for
$\ve<0.4$ whereas for large $\ve$ the separation of curves is clearly
visible. In comparison to our analytical calculations
for finite $\lambda/L$ (but still $L\to\infty$) the numerical results
show in the central region a maximum of the correlation of lower height and
position in $\ve$. It should be noted, that the
finite $\vartheta_c$ effects become more
pronounced, hence easier to observe,
 at lower CM energies, for example at $\sqrt{s}=5$ and $20$ GeV we
find $\epsilon_c=0.65$ and $\epsilon_c=0.75$ corresponding to
$\thc=10^{o}$ and $\thc=2.5^{o}$ respectively.

In summary, the angular correlations -- different from the
case of the well studied
energy distributions -- show two regimes which are dominated
either by the scaling
behaviour of the well developed cascade or the hadronisation effects
near the transverse momentum cutoff $Q_o$ respectively with specific
phenomenological consequences for both regions.

\section*{Acknowledgements}
This work is supported in part by the KBN grants PB 2P03B19609
and PB 2P30225206.

\section*{Appendix A}
In this Appendix we study in detail the analytic structure of the
Mellin representation of the two-parton density $\rho\upt$ in the constant
\al case. This serves as the basis of the generalization for the running \al
case performed Sect.III. The energy moments of the momentum distribution
\begin{equation}
\mu_n(Y)=P^{-n}\int_{Q_0/\Theta}^P k^n
\rho^{(1)}(k,P,\Theta) dk
=\int_0^Y e^{-n y} \rho^{(1)}(y,Y-y) dy, \labl{emom}
\end{equation}
read for constant \al \cite{OW3}
\begin{eqnarray}
\mu_n(Y) & = & \nonumber \\
{1\over 2\sqrt{n^2+4\a^2}  }&  ( &
      (\sqrt{n^2+4 \a^2}+n)
      \exp{[{Y\over 2}(\sqrt{n^2+4\a^2}-n)]} \nonumber    \\
  &+ & (\sqrt{n^2+4\a^2}-n)
      \exp{[-{Y\over 2}(\sqrt{n^2+4\a^2}+n)]}\;\;\;   )  \labl{muca}
\end{eqnarray}
The distribution of parent partons required in Eq.(\ref{res})
can now be reconstructed as
\begin{equation}
R(y,x)=\int_{\gamma_1} {ds \over 2\pi i} e^{s x} \mu(s,x+v)
\end{equation}
Therefore Eq.(\ref{res}) can be rewritten in the form
\begin{eqnarray}
\rho^{(2)}(\et,Y)=e^{n\a Y (1-\et)}\int_{\gamma} ds
\exp{\left((\et Y/2) (\sqrt{s^2+\a^2}+s)\right)} & & \\
\int_0^{Y(1-\et)} dy \exp{\left((y/2)(\sqrt{s^2+\a^2}-s-2n\a) \right)} &&
\end{eqnarray}
where all preexponetial factors have been omitted. The new contour
$\gamma$ follows from the change of variables $s\to -s$ and from the
interchange of the $s$ and $y$ integrations. The integration over the parent
momentum ($y=\ln(P/K)$) is now elementary and one is left with
\begin{eqnarray}
\rho^{(2)}(\et,Y)= \exp{\left(n\a Y (1-\et)\right)} & \int_{\gamma}
{ds\over 2\pi i}
\exp{\left((\et Y/2) (\sqrt{s^2+\a^2}+s)\right)} & \\
 \exp{\left(((1-\et)Y/2)(\sqrt{s^2+\a^2}-s-2n\a) \right)} &
{1\over \sqrt{s^2+\a^2}-s-2n\a} & , \labl{ana}
\end{eqnarray}
where the nonleading contribution from the lower limit have been neglected.
The final result is determined by the relative positions of the pole
and the saddle point of the integrand.
This analytical structure of is shown in
Fig.6. The pole is located at
\begin{equation}
s_{\odot}= \a {1-n^2\over n},  \labl{pole}
\end{equation}
while the saddle point position
\begin{equation}
s_{\times} = \a  {1-2\et\over \sqrt{\et(1-\et)} },\labl{sadd}
\end{equation}
depends on $\et$, and the contour $\gamma$ is now to the {\em left} of these
singularities.
For $\et<\et_c$  ( Fig.6a ) the saddle point is separated from the contour
by the
pole, and while deforming the contour
$\gamma \rightarrow \gamma' $ one picks up
the pole contribution (which is in fact dominating over the saddle
contribution). This gives the result (\ref{oc},\ref{scal1}).
For $\et > \et_c$ however $s_{\times} < s_{\odot}$  and the pole does not
contribute, see Fig.6b. In this case the saddle point saturates the integral
and one gets the $n-$independent result (\ref{escal}). Analogous structure
occurs in the running \al case as discussed in Sect. III.

 \newpage
\begin{figure}[htb]
\vspace{9pt}
\epsfxsize=10cm \epsfbox{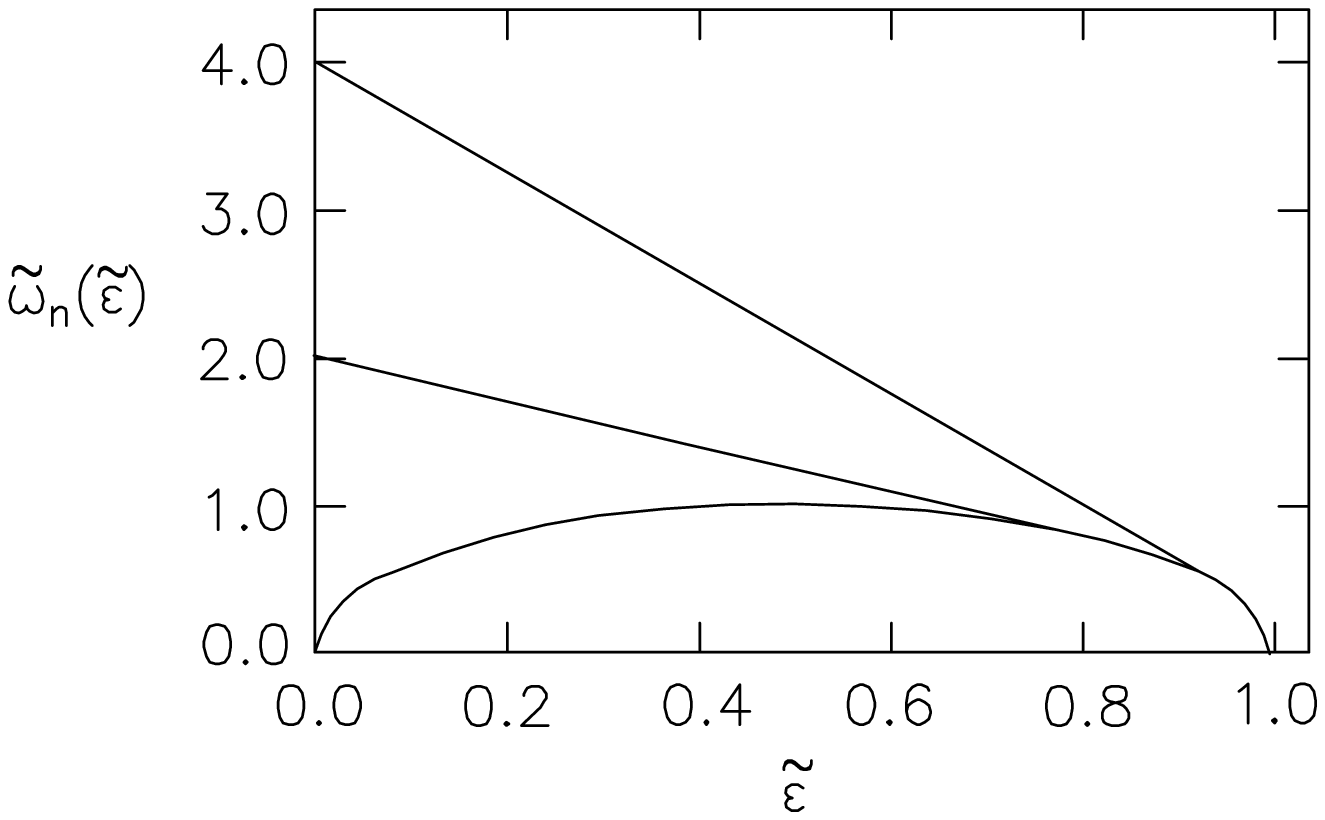}
\caption{Scaling function $\tilde{\omega}_n(\et)$ for fixed \al.
 For $\et<\et_c$ two solutions
corresponding to $n=2$ and $n=4$ are shown; for $\et>\et_c$ the solution
is given by the ellipse for all $n$.
 }
\vspace{2cm}
\label{fig:f1}
\end{figure}
\begin{figure}[htb]
\vspace{9pt}
\epsfxsize=10cm \epsfbox{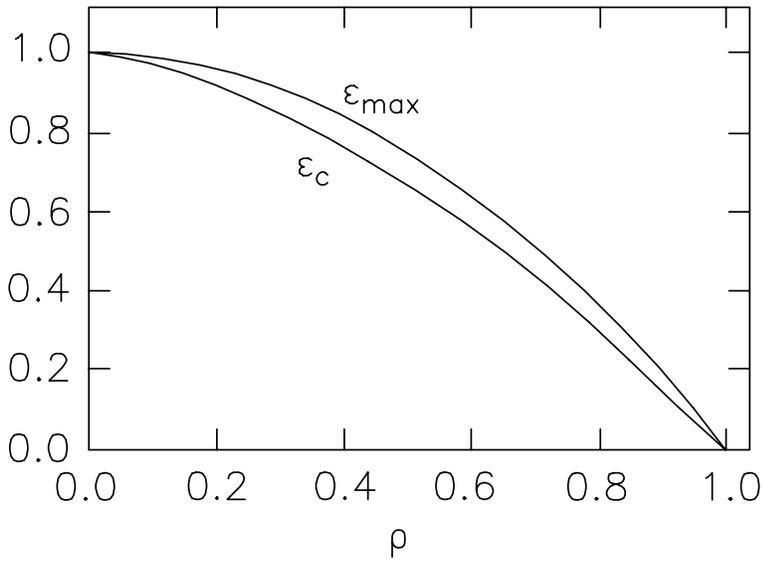}
\vspace{0.5cm}
\caption{Critical and the maximal values $\epsilon_c(\rho)$ and
$\epsilon_{max}(\rho)$ as the function of $\rho$.
 }
\label{fig:f2}
\end{figure}
\begin{figure}[htb]
\vspace{9pt}
\epsfxsize=10cm \epsfbox{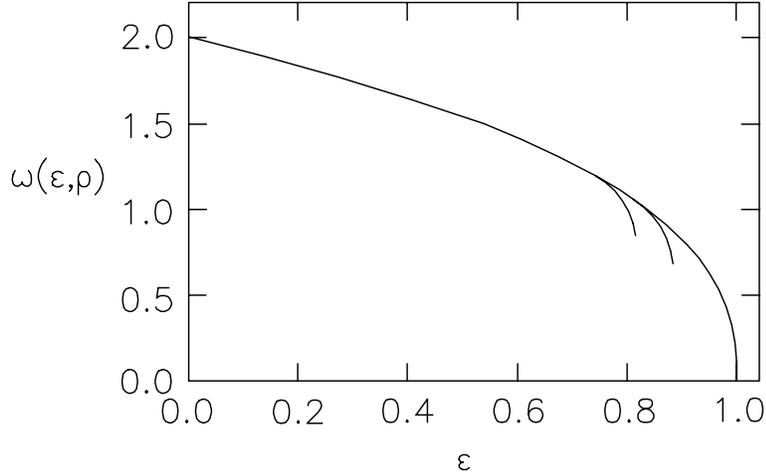}
\vspace{0.5cm}
\caption{Double scaling function $\omega_2(\epsilon,\rho)$.
The rightmost curve corresponds to the asymptotic limit $\rho=0$,
the two branches are for $\rho=0.35$ and $\rho=0.43$
(from right
to left).
 }
\vspace{0.5cm}
\label{fig:f3}
\end{figure}

\begin{figure}[htb]
\vspace{9pt}
\epsfxsize=10cm \epsfbox{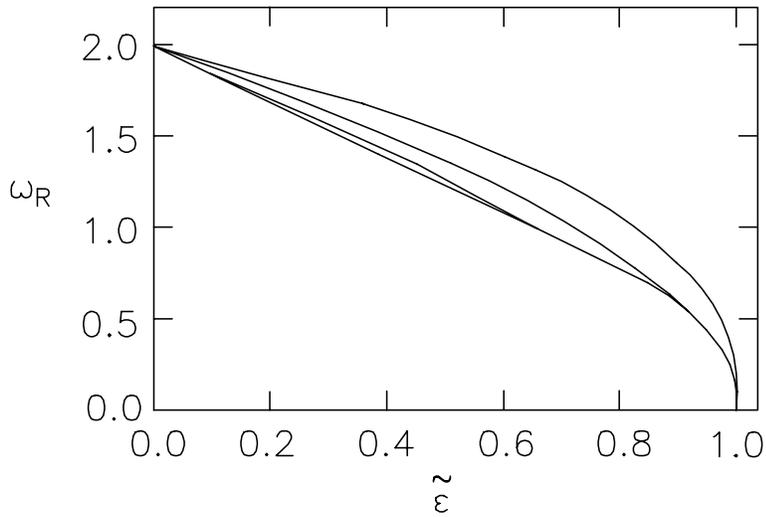}
\vspace{0.5cm}
\caption{Scaling plot of the function $\omega_R(\et,\rho)$ of
Eq.(\protect\ref{omr})
in the $\et$ variable
for $\rho=0.001,0.4,0.8,0.999$ (from top to bottom);
$\rho\sim 0 $ corresponds to the running $\alpha_s$,
 high energy limit, $\rho\sim 1$ to the constant
\al limit.
 }
\label{fig:f4}
\end{figure}
\begin{figure}[htb]
\epsfxsize=14cm
\epsfbox{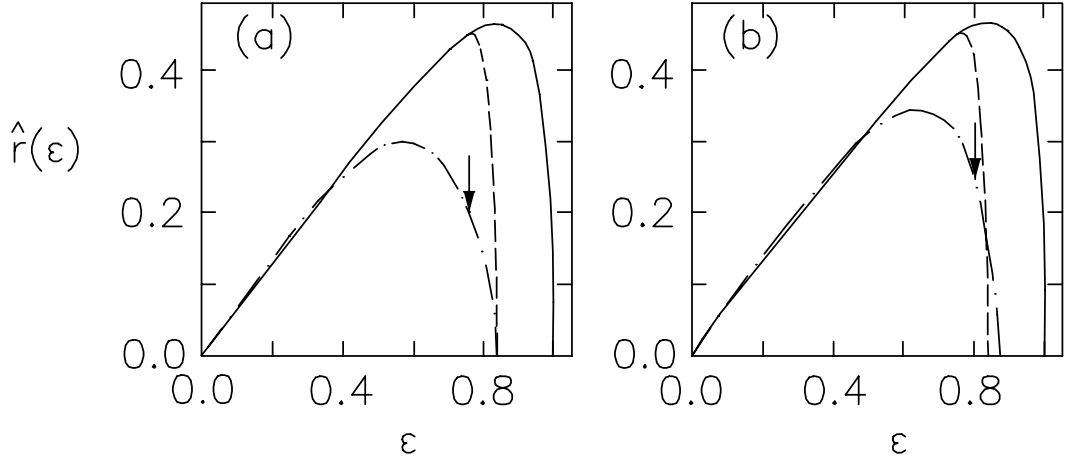}
\vspace{0.5cm}
\caption{Rescaled normalized correlation function $\hat{r}(\epsilon)$, c.f.
Eq.(\protect\ref{lnrsc})
for asymptotic energies (full lines) and for LEP energies with
jet opening angles (a) $\Theta=15^{o}$ and (b) $\Theta=60^{o}$.
The dashed curves correspond to our  asymptotic
predictions of Eq.(\protect\ref{alg},\protect\ref{algro} ),
the dash-dotted curves to the
numerical solutions
of Eq.(\protect\ref{rho2}) (for $\Lambda=0.15$ GeV, $Q_0/\Lambda=2$).
The arrow indicates the position of the critical angle}
\vspace{2cm}
\label{fig:f5}
\end{figure}

\begin{figure}[htb]
\epsfxsize=14cm
\epsfbox{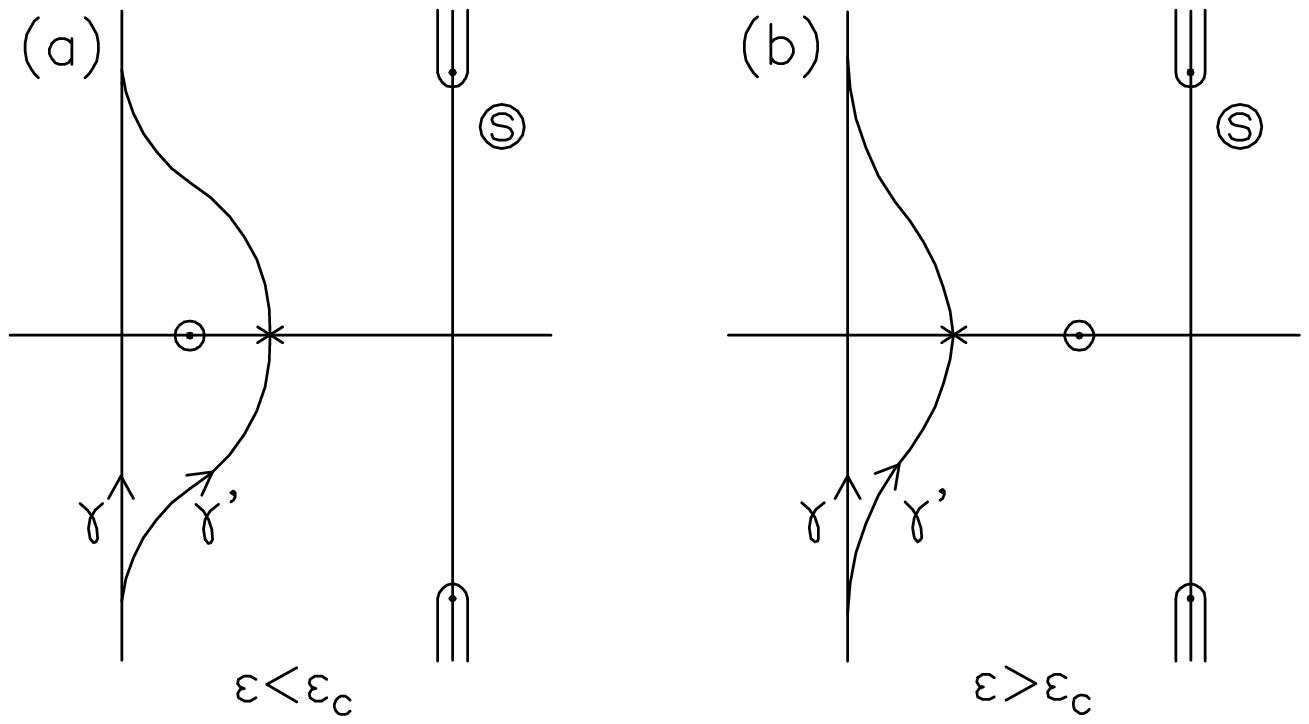}
\vspace{0.5cm}
\caption{Analytical structure of the Mellin representation
Eq.(\protect\ref{ana}) in
the complex $s$ plane; (a) $\epsilon < \epsilon_c$ and (b) $\epsilon >
\epsilon_c$.
    }
\label{fig:f6}
\end{figure}



\begin{thebibliography}{99}
\bibitem{DMO} Yu. L. Dokshitzer, G. Marchesini, G. Oriani,
\np{B387}{1992}{675}.
\bibitem{OW1} W. Ochs, J. Wosiek, \pl{B289}{1992}{159}.
\bibitem{OW2} W. Ochs, J. Wosiek,
\pl{B304}{1993}{144};~ \zp{C68}{1995}{269},~ see~also
{}~preprint~hep-ph/9412384.
\bibitem{DD} Yu. L. Dokshitzer, I. Dremin, \np{B402}{1993}{139}.
\bibitem{BMP} Ph. Brax, J.L. Meunier, R. Peschanski,
\zp{C62}{1994}{649}.
\bibitem{LPHD} Ya.I. Azimov, Yu.L. Dokshitzer, V.A. Khoze, S.I. Troyan,
\zp{C27}{1985}{65};~ \zp{C31}{1961}{21}.
\bibitem{DELPHI} F. Mandl, B. Buschbeck (DELPHI coll.), Proc. XXIV
  Int. Symp. on Multiparticle Dynamics, Vietri sul mare, 1994,
 Eds. A. Giovannini, S. Lupia, R. Ugoccioni, World Scientific,
Singapore (1995) p.52.
\bibitem{EMMC} Yu. Dokshitzer, V.A. Khoze, G. Marchesini, B.R. Webber,
\pl{B245}{1990}{243}.
\bibitem{OW3} W. Ochs, J. Wosiek, Proc. Int. Europhysics Conf.
  on High Energy Physics, Brussels, July 1995, to be publ. by World
Scientific, Singapore.


\end{thebibliography}
 \end{document}